\documentclass[conference]{IEEEtran1}
\IEEEoverridecommandlockouts

\usepackage{lineno,hyperref}
\usepackage{graphicx}
\usepackage{subcaption}
\usepackage{textcomp}
\usepackage{xcolor}
\usepackage{multirow}
\usepackage{booktabs}
\usepackage{makecell}
\usepackage{stackengine}
\usepackage{longtable}
\usepackage{amsmath,amssymb,amsfonts}
\usepackage{cleveref}
\usepackage{placeins}
\usepackage{listings}
\usepackage{xspace}
\usepackage{comment}
\usepackage{acronym}

\usepackage{float}
\usepackage{needspace}
\usepackage{cite}
\newlength{\commentindent}
\usepackage{algorithm,algorithmicx,algpseudocode}
\usepackage{listings}
\usepackage{xcolor}

\definecolor{codegreen}{rgb}{0,0.6,0}
\definecolor{codegray}{rgb}{0.5,0.5,0.5}
\definecolor{codepurple}{rgb}{0.58,0,0.82}
\definecolor{backcolour}{rgb}{0.95,0.95,0.92}

\lstdefinestyle{mystyle}{
    commentstyle=\color{codegreen},
    keywordstyle=\color{magenta},
    numberstyle=\tiny\color{codegray},
    stringstyle=\color{codepurple},
    basicstyle=\ttfamily\footnotesize,
    breakatwhitespace=false,         
    breaklines=true,                 
    captionpos=b,                    
    keepspaces=true,                 
    showspaces=false,                
    showstringspaces=false,
    showtabs=false,                  
    tabsize=2
}

\acrodef{acopf}[ACOPF]{alternating current optimal power flow}
\acrodef{cpu}[CPU]{central processing unit}
\acrodefplural{cpu}[CPUs]{central processing units}
\acrodef{gpu}[GPU]{graphical processing unit}
\acrodefplural{gpu}[GPUs]{graphical processing units}
\acrodef{simd}[SIMD]{single-instruction multiple-data}

\newcommand{\cusolver}{cuSOLVER\xspace}
\newcommand{\cusolverglu}{\texttt{cusolverGLU}\xspace}

\newcommand{\nvidia}{NVIDIA\xspace}
\newcommand{\hiop}{HiOp\xspace}
\newcommand{\exago}{ExaGO\textsuperscript{TM}\xspace}

\begin{document}

\title{Towards Efficient Alternating Current  Optimal Power Flow Analysis on Graphical Processing Units
\thanks{
This research was supported by the Exascale Computing Project (17-SC-20-SC), a collaborative effort of the U.S. Department of Energy Office of Science and the National Nuclear Security Administration. This research used resources of the Oak Ridge Leadership Computing Facility, which is supported by the U.S. Department of Energy Office of Science under Contract No. DE-AC05-00OR22725.
}
}
\author{
\IEEEauthorblockN{Kasia \'{S}wirydowicz}
\IEEEauthorblockA{\textit{Advanced Computing, Mathematics and Data Division} \\
\textit{Pacific Northwest National Laboratory}\\
Richland, WA, United States \\
kasia.swirydowicz@pnnl.gov}
\and
\IEEEauthorblockN{Nicholson Koukpaizan}
\IEEEauthorblockA{\textit{National Center for Computational Sciences} \\
\textit{Oak Ridge National Laboratory}\\
Oak Ridge, TN, United States \\
koukpaizannk@ornl.gov}
\and
\IEEEauthorblockN{Shrirang Abhyankar}
\IEEEauthorblockA{\textit{Electricity Infrastructure and Buildings Division} \\
\textit{Pacific Northwest National Laboratory}\\
Richland, WA, United States \\
shri@pnnl.gov}
\and
\IEEEauthorblockN{Slaven Pele\v{s}}
\IEEEauthorblockA{\textit{Computational Sciences and Engineering Division} \\
\textit{Oak Ridge National Laboratory}\\
Oak Ridge, TN, United States \\
peless@ornl.gov}
}

\maketitle

\begin{abstract}
We present a solution of sparse \ac{acopf}
analysis on \ac{gpu}. In particular, we discuss the performance bottlenecks and detail our efforts to accelerate the linear solver, a core component of ACOPF that dominates the computational time. ACOPF analyses of two large-scale systems, synthetic Northeast (25,000 buses) and Eastern (70,000 buses) U.S. grids \cite{birchfield2017tamu-cases}, on \ac{gpu} show promising speed-up compared to analyses on \ac{cpu} using a state-of-the-art solver. To our knowledge, this is the first result demonstrating a significant acceleration of sparse \ac{acopf} on \acp{gpu}.
\end{abstract}


\IEEEpeerreviewmaketitle

\section{Introduction}

The electric power grid is undergoing a massive transformation with the influx of new technologies, processes, and digitization to provide cleaner, smarter, affordable, and equitable electricity to everyone. Yet, at the same time, the grid is witnessing an unprecedented increase in threats caused by climate change and cyber attacks. Blackouts caused in California due to heat waves, outages due to hurricanes in Florida and Puerto Rico, are some recent examples of massive power disruptions affecting millions of lives. Operating the grid under such ``dark sky'' events and, in general, ``intermittent sky'' days with uncertain renewable forecasts, remains an ever-increasing challenge. Accurate and fast wide area analysis tools provide critical aid to grid operators and planners, as they make decisions. However, with greater fidelity comes greater computational complexity.

Heterogeneous computing, such as combining \acp{cpu} with \emph{hardware accelerators}, is becoming a dominant paradigm in the computing landscape. Spurred by demand from the video gaming industry, artificial intelligence and computer vision applications, \acp{gpu} are the most prevalent hardware accelerators today. \ac{gpu}s deliver high computational power at low cost through massive fine-grain parallelism.

The challenge for using \ac{gpu}s arises from their \ac{simd} architecture that imposes modeling and solver constraints. The \ac{simd} architecture is suitable for porting legacy CPU-based \textit{dense} vector and matrix operations. Legacy \textit{sparse} operations, which are used in current state-of-the-art \ac{acopf} analysis \cite{ONeill2012} cannot be recast in \ac{simd} terms without additional mathematical considerations.

In our prior work \cite{abhyankar2021acopf}, we discussed solving \ac{acopf} \cite{ONeill2012, Frank2012} on \ac{gpu}s with a dense solver, by compressing the sparse formulation that naturally arises from power systems. Earlier, an approach for solving convex \ac{acopf} problems using fully dense linear algebra and Cholesky solver was proposed in \cite{rakai2014gpu}. However, the computational complexity of dense linear solvers is $O(N^3)$ and their memory complexity is $O(N^2)$, where $N$ is the number of linear equations. For larger grid models, all performance gains on \ac{gpu} are offset by a cubic increase in computational cost and a quadratic increase in memory.


Here, we present our results on solving \emph{sparse} \ac{acopf} on \ac{gpu}s. In particular, we detail our efforts on accelerating the linear solver, a core component of \ac{acopf} that dominates the computational cost \cite{su2019full}. This exploration is motivated by our recent efforts on sparse linear solvers for \ac{acopf} problems \cite{swirydowicz2022linear, regev2022kkt}, as well as recent investigations of dynamic phasor simulations \cite{dinkelbach2021factorisation, razik2019comparative}, power flow  \cite{ablakovic2012parallel,dorto2021comparing}, and batched power flow simulations \cite{zhou2017gpu,zhou2018gpu} on \acp{gpu}. 

The main contributions of this paper are:
\begin{itemize}
    \item First reported meaningful acceleration of \ac{acopf} on \ac{gpu}s, using our \emph{sparse} linear solver, demonstrated on large interconnection-level grid models and reproducible using the open-source \exago package (Sec.~\ref{sec:gpu}).
    \item Performance analysis of \ac{acopf} on \ac{cpu}s (Sec.~\ref{sec:opf}) and \ac{gpu}s (Sec.~\ref{sec:gpu}) with key bottlenecks identified.
    \item Performance projections and limitations of current technology and identifying research needed to achieve even better acceleration on \ac{gpu}s (Sec.~\ref{sec:projection} and~\ref{sec:conslusion}). 
\end{itemize}


\section{State of The Art Methods for Alternating Current Optimal Power Flow Analysis}\label{sec:opf}

\ac{acopf} falls in the category of nonlinear nonconvex optimization problems. In compact mathematical form it can be formulated as
\begin{align}
\underset{x}{\min}&~ F(x)& \label{eq:opfobj}\\
\text{s.t.}~~     & g(x) = 0        & \label{eq:opfeq}\\
h^{-}& \le h(x) \le h^{+}& \label{eq:opfineq}\\
x^{-}& \le x \le x^{+} & \label{eq:opfbounds}
\end{align}
where $F(x)$ is a scalar function defining the optimization objective, typically the total cost of generation; $x$ is the vector of optimization variables, such as generator power outputs; 
vectors $x^{-},x^{+}$ define variable bounds, e.g.~resource capacity limits; vector function $g(x)$ defines equality constraints, like power balance; and vector function $h(x)$ defines security constraints with security limits given by vectors $h^{-},h^{+}$. The reader is referred to \cite{Zimmerman2011,ONeill2012} for comprehensive details on the \ac{acopf} formulation.


The optimization problem (\ref{eq:opfobj})-(\ref{eq:opfbounds}) is typically solved in an iterative manner  using an interior-point method \cite{wachter2006implementation}. To keep our presentation streamlined and without loss of generality, we rewrite (\ref{eq:opfobj})-(\ref{eq:opfbounds}) in a more compact form
\begin{align}
\underset{y}{\min}~& f(y)& \label{eq:opfcompactobj}\\
\text{s.t.}~~& c(y) = 0& \label{eq:opfcompacteq}\\
             & y \ge 0& \label{eq:opfcompactbounds}
\end{align}
Here we convert (\ref{eq:opfineq}) to equality constraints 
by adding slack variables $s',s'' \ge 0$ satisfying $h(x)-h^- - s'$ and $h(x)-h^+ + s''$. Bounds (\ref{eq:opfbounds}) are expressed in form (\ref{eq:opfcompactbounds}) by introducing variables $x',x'' \ge 0$ such that $x - x^- = x'$ and $x^+ - x = x''$. In the compact form, the vector of optimization primal variables is $y=(x',x'',s',s'') \in \mathbb{R}^n$ and $f: \mathbb{R}^n \to \mathbb{R}$ is the scalar objective function $F(x)$ expressed in space of primal variables $y$. All constraints are formulated in $c: \mathbb{R}^n \to \mathbb{R}^m$, where $m$ is the number of equality and inequality constraints.

To enforce variable bounds, the objective function (\ref{eq:opfcompactobj}) is augmented with barrier functions as
\begin{equation}
 \phi(y) = f(y) + \mu \sum_{i=1}^n \ln y_i.
\end{equation}
The solution to the barrier subproblem then can be obtained by solving the nonlinear equations
\begin{align}
 \nabla{f(y)} + \nabla{c(y)}\lambda - z &=0, \label{eq:gradient}\\
     c(y) & = 0, \label{eq:constraint} \\
     Yz - \mu e & = 0 \label{eq:bounds}
\end{align}
where $J\equiv \nabla{c(y)}$ is the $m \times n$ constraint Jacobian matrix, $\lambda \in \mathbb{R}^m$ and $z \in \mathbb{R}^n$ are vectors of Lagrange multipliers for equality (\ref{eq:constraint}) and bound constraints (\ref{eq:bounds}), respectively, $Y \equiv \text{diag}(y)$, and $e \in \mathbb{R}^n$ is a vector of ones.

The barrier subproblem (\ref{eq:gradient})-(\ref{eq:bounds}) is solved using a Newton method, and the solution to the original optimization problem is obtained by a continuation algorithm setting $\mu \to 0$. As it searches for the optimal solution, the interior-point method solves a sequence of linearized systems $K_k \Delta x_k = r_k$, $k=1,2,\dots, M$ of Karush-Kuhn-Tucker (KKT) type. A typical implementation of an interior-point method eliminates $z$ from the linearized version of (\ref{eq:gradient})-(\ref{eq:bounds}) by block-Gaussian elimination to obtain smaller symmetric system
\begin{align} \label{eq:kktlinear}
\overbrace{\begin{bmatrix}
      H + D_y & J
     \\ J^T   & 0 
  \end{bmatrix}}^{K_k}
  \overbrace{\begin{bmatrix}
    \Delta y \\ 
    \Delta \lambda
  \end{bmatrix}}^{\Delta x_k}=
  \overbrace{\begin{bmatrix}
    r_{y} \\ r_{\lambda}
  \end{bmatrix}}^{r_k},
\end{align}
where index $k$ denotes the optimization solver iteration (including continuation step in $\mu$ and Newton iterations), $K_k$ is a system (KKT) matrix, vector $\Delta x_k$ is a search direction for the primal and dual variables, and $r_k$ is derived from the residual vector for (\ref{eq:gradient})-(\ref{eq:bounds}) evaluated at the current value of the primal and dual variables.
%
%
The Hessian
\begin{equation*}
H \equiv \nabla^2 f(y) + \sum_{i=1}^{m} \lambda_{i} \nabla^2 c_i(y),
\end{equation*}
is a sparse symmetric $n \times n$ matrix and 
$D_y \equiv \mu Y^{-2}$ is a diagonal  $n \times n$ matrix. 
For more details on interior-point methods implementations we refer reader to \cite{wachter2006implementation}.

\begin{figure}[htb]
\centering
  \includegraphics[width=\columnwidth]{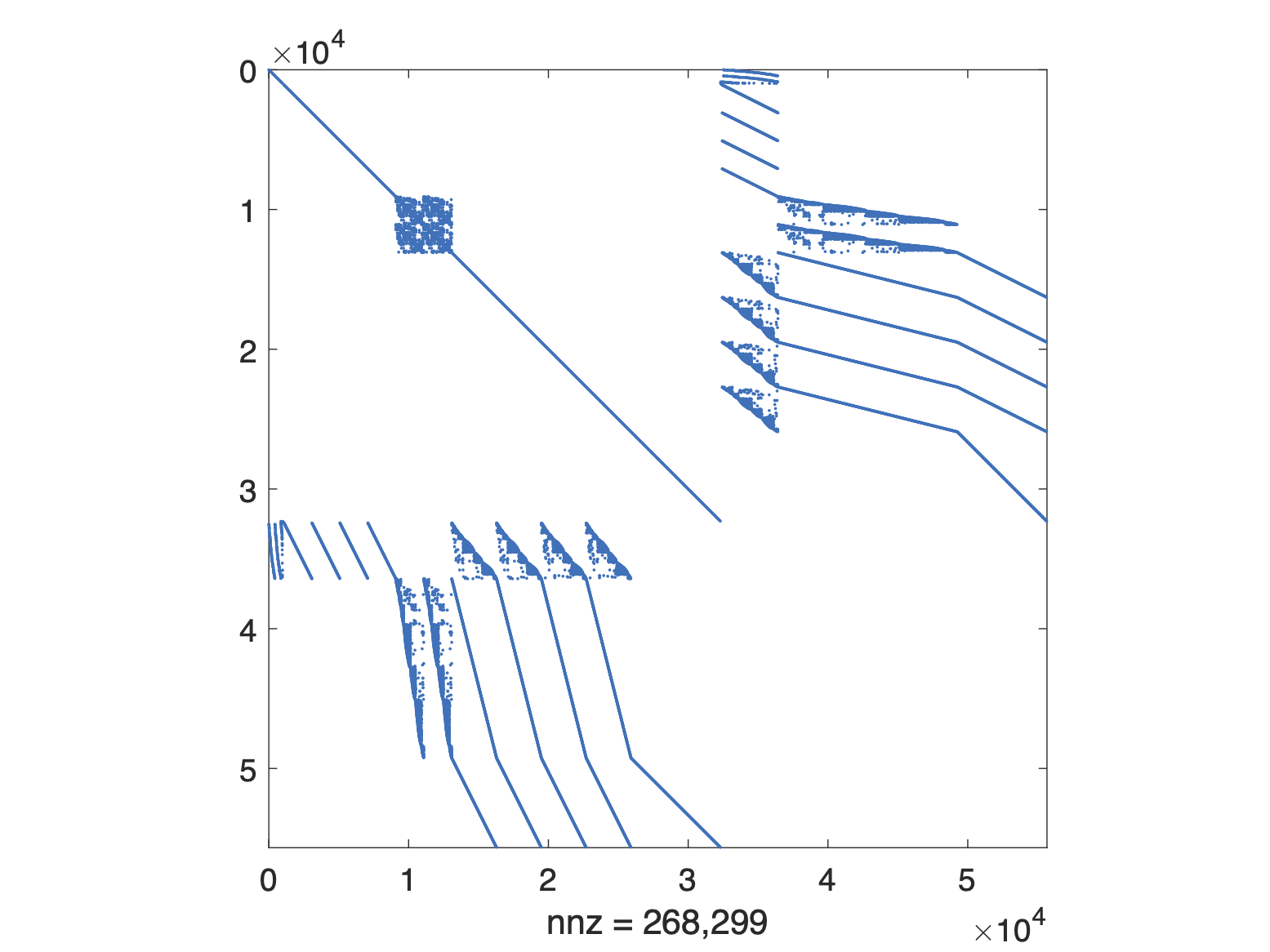}
  \caption{Sparsity pattern of KKT matrix obtained in \ac{acopf} analysis for 2000-bus Texas grid model.
  }
  \label{fig:sparsity}
\end{figure}

Matrices $K_k$ in (\ref{eq:kktlinear}) are sparse symmetric indefinite. All $K_k$ have the \textit{same sparsity pattern}, a property that can be exploited by a solver. Characteristic for ACOPF is that $K_k$ generated in the analysis are \textit{very sparse with an irregular non-zero pattern} without dense blocks \cite{deAraujo2013}. One such example is shown in Figure \ref{fig:sparsity}.

When the barrier parameter $\mu=0$, the matrix in (\ref{eq:kktlinear}) is singular. The continuation algorithm driving $\mu \to 0$ therefore needs to exit when the solution to the augmented system is close enough to the solution of the original system, but before (\ref{eq:kktlinear}) becomes too ill-conditioned. Typically, KKT matrices are characterized by \textit{extremely high condition numbers} ($>10^{16}$), which means that solving these systems in a naive way (e.g. without pivoting and equilibration) would result in solutions with error exceeding computer floating point double precision.

Only a few linear solvers have been effective for this type of problems. These solvers are typically sparse direct solvers based on $\text{LDL}^\text{T}$ or $LU$ decomposition \cite{golub2013matrix}. State-of-the art approaches usually employ KLU (for non-symmetric matrices) \cite{davis2010algorithm, razik2019comparative} or MA57~\cite{duff2004ma57}. 

A sparse direct solver typically computes the solution to a linear system in three stages: ($i$) an analysis stage, where the system is equilibrated, column and row permutations are set, pivot order is established and the non-zero pattern of the factors is chosen; ($ii$) a numerical factorization stage, in which matrix factors are computed and ($iii$) a solution stage, in which the factors computed in the previous stage are used for forward and backward triangular solves. Some solvers follow the third stage by iterative refinement~\cite{skeel1980iterative}. The first stage is considered to be more expensive than others, but if the matrix non-zero pattern does not change between the systems, as in (\ref{eq:kktlinear}), the analysis can be executed once and the results reused for all $K_k$ \cite{razik2019comparative,dorto2021comparing}. The numerical factorization stage typically requires pivoting for the computation to remain stable. Pivoting changes the sparsity pattern of $L$ and $U$ factors and in some cases may even require memory reallocation. Therefore, while changes in the pivot sequence may not be a concern on \acp{cpu}, they degrade performance on \acp{gpu} as we describe in Sec.~\ref{sec:gpu}.

\begin{figure}[h!]
\centering
  {\includegraphics[width=0.8\columnwidth,trim={1.1cm 0.4cm 1.1cm 7.5cm},clip]{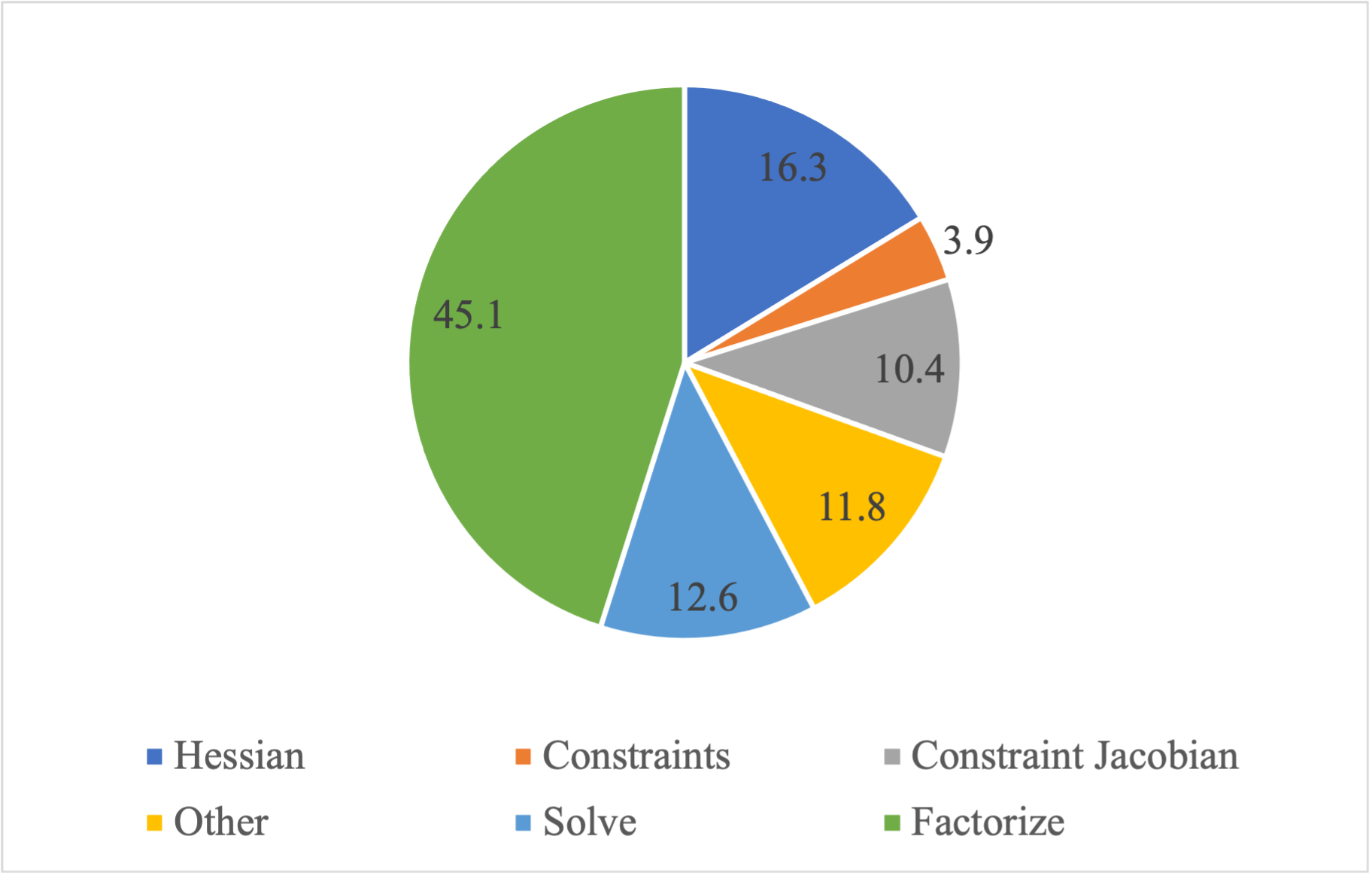}}
  \begin{subfigure}{0.49\columnwidth}
    {\includegraphics[width=\columnwidth,trim={3cm 2cm 3cm 0.1cm},clip]{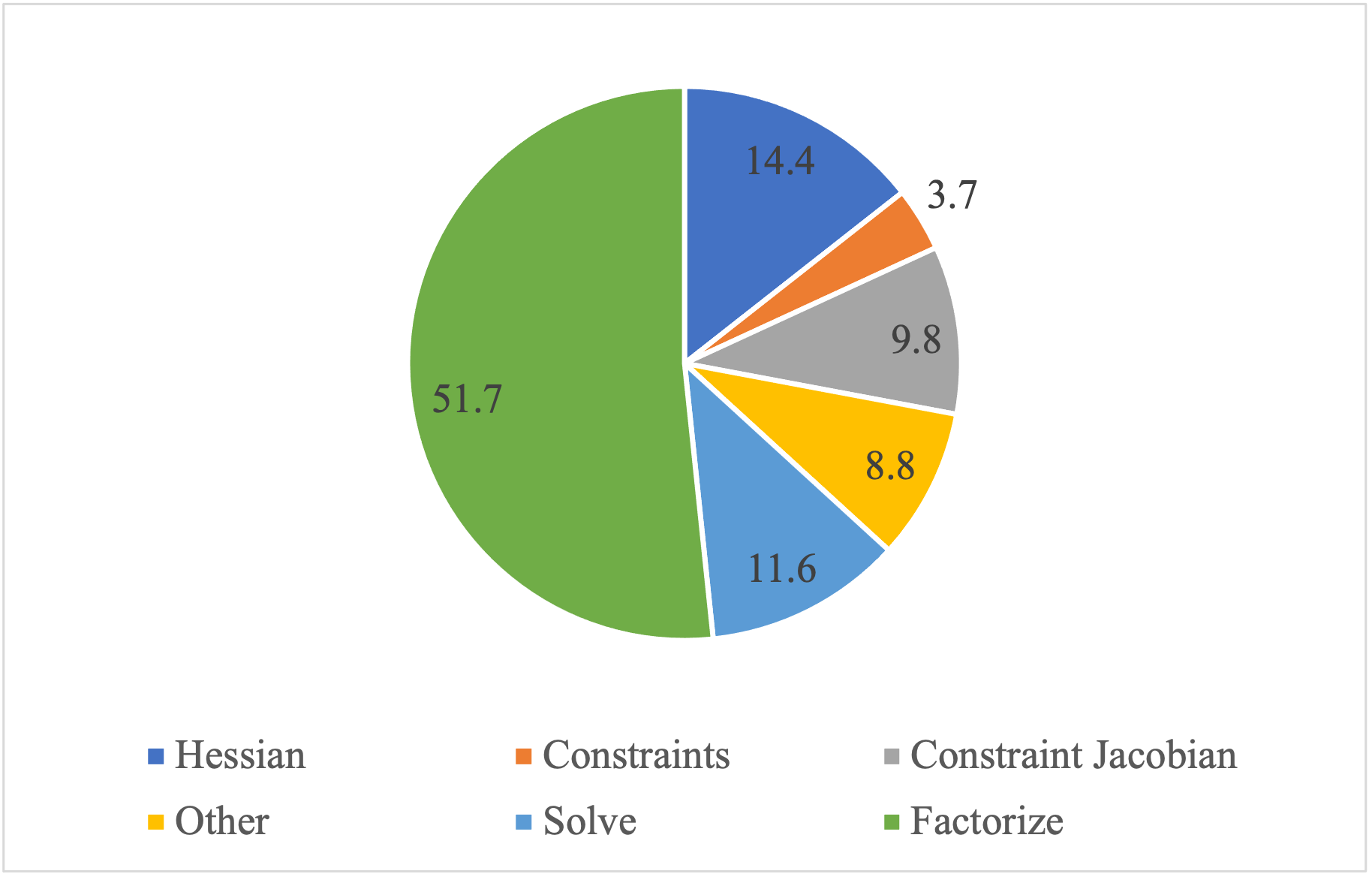}}
    \caption{Northeast U.S. grid}
  \end{subfigure}
  \begin{subfigure}{0.49\columnwidth}
    {\includegraphics[width=\columnwidth,trim={3cm 2cm 3cm 0.1cm},clip]{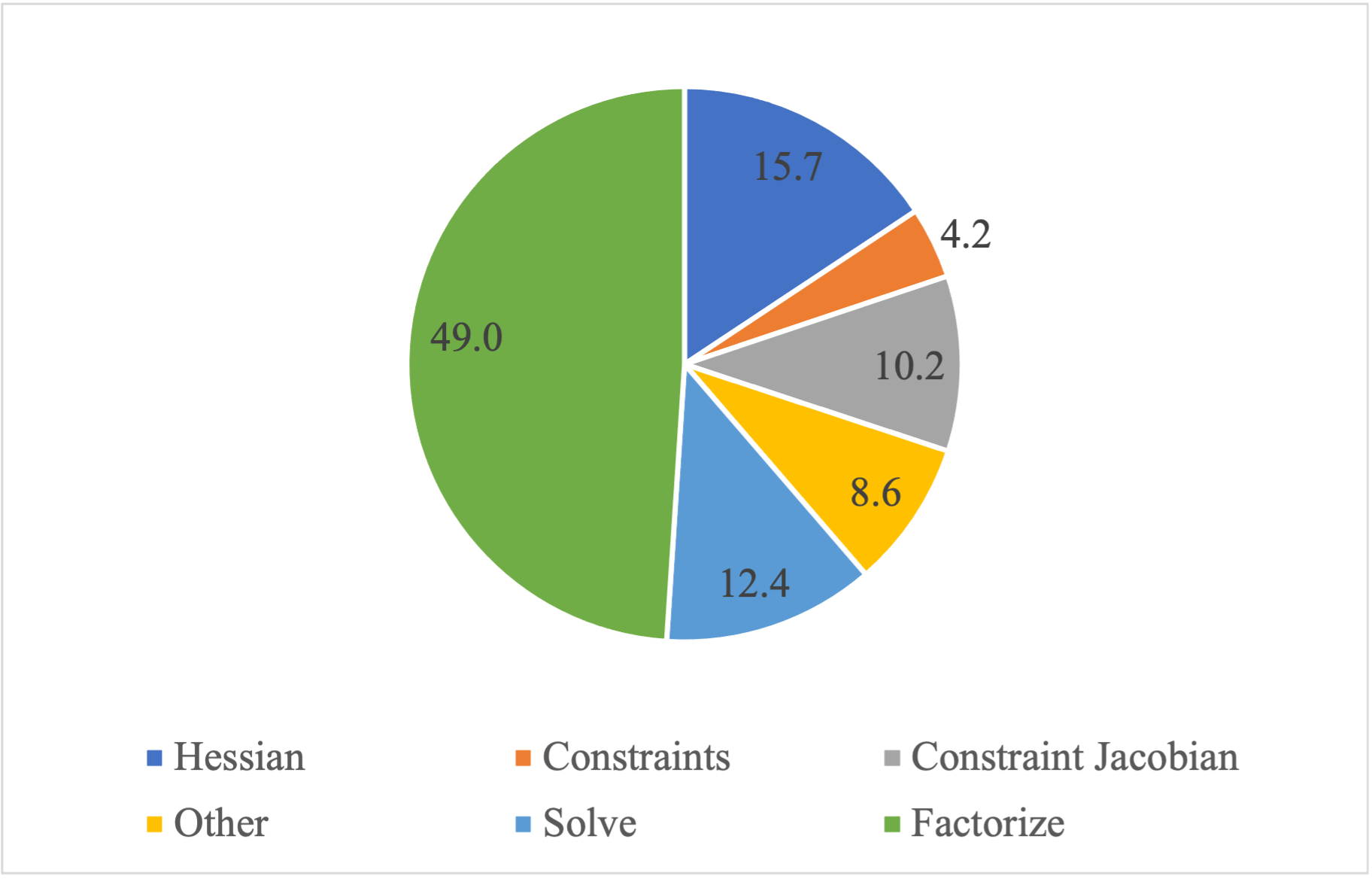}}
    \caption{Eastern U.S. grid}
  \end{subfigure}

  \caption{Computational cost of ACOPF broken down by solver functions. Linear solver functions (matrix factorization and triangular solve) contribute to 60\% of the overall ACOPF compute time.}
  \label{fig:ma57cost}
\end{figure}

For \ac{acopf} problems, more than a half of the overall computational cost is typically spent solving the KKT linear system. Within that, the largest cost is in matrix factorization (symbolic and numeric factorization) and to a lesser degree in triangular solves (backward-forward substitution). Other significant items are evaluation of derivative matrices -- constraint Jacobian $J$, and Hessian matrix $H$. We illustrate that in Fig. \ref{fig:ma57cost} where we show computational cost breakdown for \ac{acopf} for 25,000-bus model of Northeast U.S. grid and 70,000-bus model of Eastern U.S. grid, respectively. Table \ref{tab:tamu_cases} provides a description of the two synthetic grids evaluated.

\begin{table}[h!]
    \centering
     \caption{Description of synthetic grid models from \cite{birchfield2017tamu-cases} used in performance analysis. The specifics of the linear system (\ref{eq:kktlinear}) for each of these models are given in terms of the matrix $K_k$ size (N) and number of non-zeros (nnz). }
    \begin{tabular}{|l|c|r|c|c|c|}
    \hline
    \textbf{Grid} & \textbf{Buses} & \textbf{Gen's} & \textbf{Branches} & \textbf{N} & \textbf{nnz}  \\ \hline
     Northeast U.S   & 25,000 & 4,834 & 32,230 & 107,558 & 1,191,152 \\ \hline
     Eastern U.S.    & 70,000 & 10,390 & 88,270 & 296,214 & 3,228,456 \\ \hline
    \end{tabular}
   
    \label{tab:tamu_cases}
\end{table}

The analysis has been performed using the \exago package \cite{ExaGo} with the \hiop optimization engine \cite{hiop_techrep} and the MA57 linear solver. Numerical matrix factorization takes roughly half of the total compute time in \ac{acopf}, while the triangular solve takes another 12\%. Model evaluation operations take about 30\% of the total compute time. All other operations (evaluation of the objective, objective gradient, infeasibility norm, etc.) in \ac{acopf} take less than 10\% of the overall time.
Achieving any speedup largely depends on how effectively the KKT linear system can be solved on a \ac{gpu}.

\section{Computing with graphical processing units}
\label{sec:gpu}

There has been a substantial effort in the scientific community devoted towards using \ac{gpu} acceleration to speed up sparse direct linear solvers, and the available software usually has at least some \ac{gpu} capabilities. 
However, solving the linear systems described herein on \acp{gpu} is particularly challenging: ($i$) Due to irregular and very sparse nature of power systems, the underlying linear problems do not have a structure that traditional multifrontal~\cite{duff1983multifrontal} and supernodal~\cite{demmel1999supernodal} approaches can use to form dense blocks where \ac{simd} operations can be performed. ($ii$) The underlying linear problems are ill-conditioned and require pivoting for numerical factorization stability. Pivoting degrades performance on \acp{gpu} because of conditional branching and considerable data movement \cite{zhou2017gpu}.


In \cite{swirydowicz2022linear}, we identified and tested several \ac{gpu} accelerated sparse direct solvers and concluded that none of them was faster than MA57, especially for the largest test cases. 
These solvers only offload a part of the computations (e.g., matrix-matrix products or triangular solves) to the \ac{gpu}, while the rest of the computation happens on the \ac{cpu}. This results in a substantial overhead associated with allocating and freeing small chunks of \ac{gpu} memory and copying the data to and from the device. For our test cases, we also observed that factorization on the \ac{gpu} was particularly expensive. 

We use a refactorization approach to solve (\ref{eq:kktlinear}) on \ac{gpu} (see Algorithm~\ref{alg:kluglu}, $M$ is the total number of systems to solve, this number depends on optimization solver settings, typically in the order of 100s). A similar approach was successfully used for dynamic phasor \cite{razik2019comparative,dinkelbach2021factorisation} and power flow \cite{dorto2021comparing} simulations but not for \ac{acopf}. Our approach is the first we are aware of, where the sparse linear solvers in ACOPF analysis were accelerated using GPUs and outperformed state-of-the-art CPU solvers.  We replace the MA57 solver in our software stack with a solver we implemented using KLU and NVIDIA's \texttt{cuSolver} libraries. We factorize the first system on the \ac{cpu} using KLU, set the patterns of $L$ and $U$, and compute the permutation vectors. We also evaluated \texttt{cuSolver}'s methods for setting the patterns of $L$ and $U$, but found KLU to perform better. Once computed, we keep the sparsity pattern of the factors the same for the rest of the computation. The non-zero structure of the factors and the permutation vectors are then copied to the \ac{gpu} and passed to \texttt{cuSolver}. The refactorization is set up, and each subsequent system is solved using \ac{gpu} refactorization, without a need for pivoting, which is prohibitively expensive on \ac{gpu}s. 

\begin{algorithm*}[hbt]
\caption{Refactorization Solution Strategy: KLU + \texttt{cuSolverGLU}}\label{alg:kluglu}
 \hspace*{\algorithmicindent} \textbf{Input:} Sequence of linear systems $K_kx_k=b_k$, $k=1, 2, \ldots, M$ \\
 \hspace*{\algorithmicindent} \textbf{Output:} Sequence of solution vectors $x_k$
\begin{algorithmic}[1]
\State{Use KLU to solve $Ax_1=b_1$}
\State{Extract symbolic factorization and permutation vectors. \Comment{Using \texttt{klu\_extract}, returns L,U in CSC format}}
\State{Convert CSC to combined L+U CSR object.} \Comment{Up to here, computed on the \ac{cpu}}
\State{Setup \texttt{cuSolverGLU}}\Comment{\ac{gpu} computation starts}
\For{$k=2,\ldots, M$} \Comment{$M$ is the total number of linear systems}
\State Update factorization object using new values from $K_k$.~~~~ \Comment{Using \texttt{cuSolverGLU} function}
\State Refactor. 
\State Perform triangular solve for $b_k$. \Comment{Always followed by built-in iterative refinement}
\EndFor
\end{algorithmic}
\end{algorithm*}

Using this method all data movement and mapping is done only once on the \ac{cpu} while the subsequent computations are implemented in terms of \ac{simd} operations on \ac{gpu}.
The tradeoff is that the refactorization approach produces a lower quality solution, so the solve stage may need to be followed by an iterative refinement to recover the accuracy needed by the optimization method.

Two refactorization interfaces were tested: \texttt{cuSolverGLU} and \texttt{cuSolverRf}. \texttt{cuSolverGLU} is an undocumented (but publicly available) part of the \texttt{cuSolver} library. It is generally faster than \texttt{cuSolverRf}, more numerically stable and comes with built-in iterative refinement. We used \texttt{cuSolverGLU} for all the results shown in this paper. \texttt{cuSolverGLU} requires a combined $L+U$ storage of the factors in CSR format hence we needed to perform a format conversion after using KLU, which operates on matrices in CSC format.


A great advantage of \texttt{cuSolver} refactorization is that the user is allowed to provide the non-zero patterns of their own $L$ and $U$ factors and the permutation vectors; the factors can come from \textit{any} $LU$ solver (as long as the data format is correct) and \texttt{cuSolver} will carry on the computation on the \ac{gpu} for the next systems in the sequence. 

\begin{figure}[htb]
\centering
  {\includegraphics[width=0.8\columnwidth,trim={1.1cm 0.4cm 1.1cm 7.5cm},clip]{./ma57.25k.pie}}
  \begin{subfigure}{\columnwidth}
    {\includegraphics[width=\columnwidth,trim={0.1cm 1cm 0.1cm 0.1cm},clip]{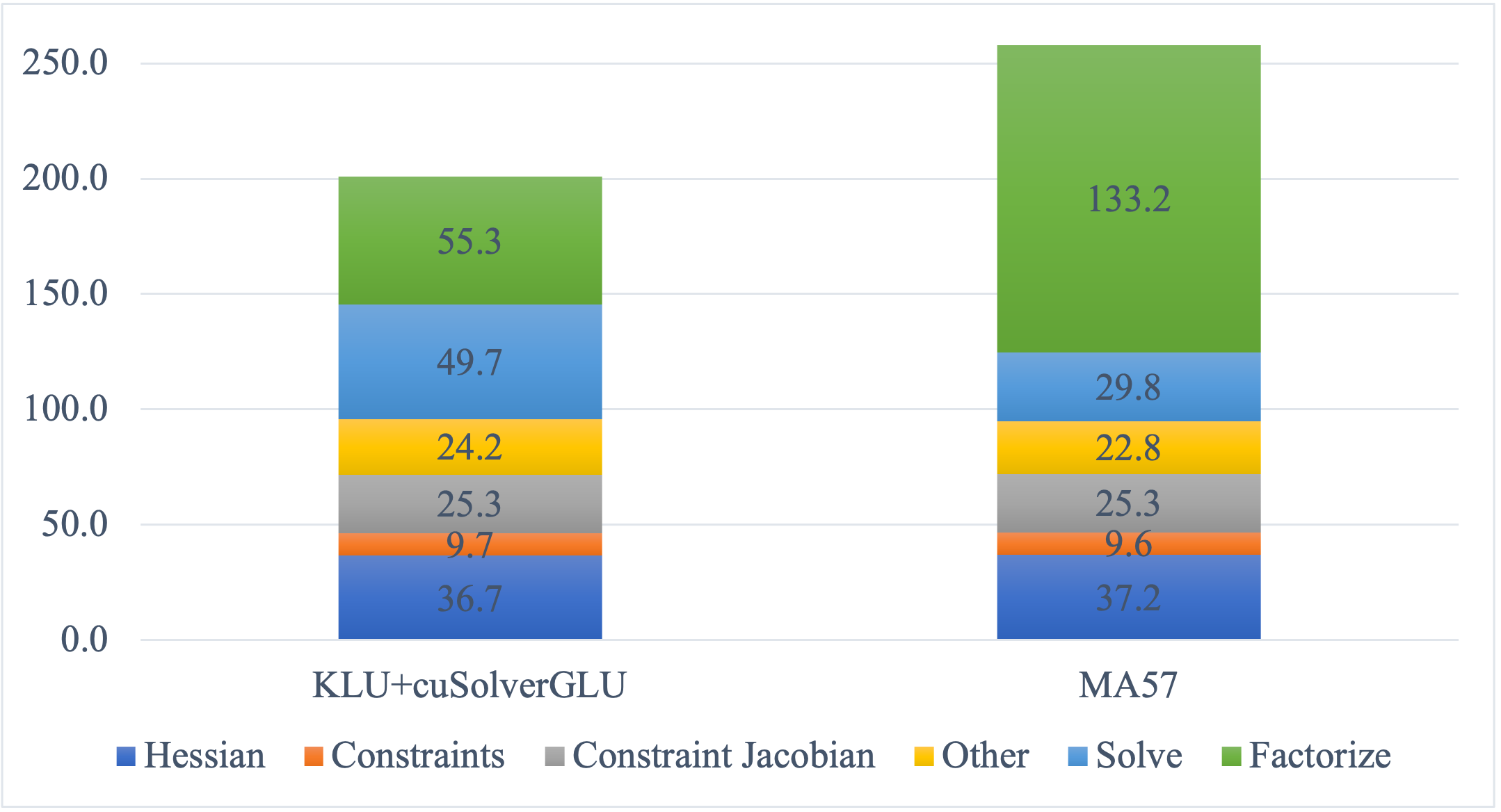}}
    \caption{Northeast U.S. grid}
  \end{subfigure}
  \begin{subfigure}{\columnwidth}
    {\includegraphics[width=\columnwidth,trim={0.1cm 1cm 0.1cm 0.1cm},clip]{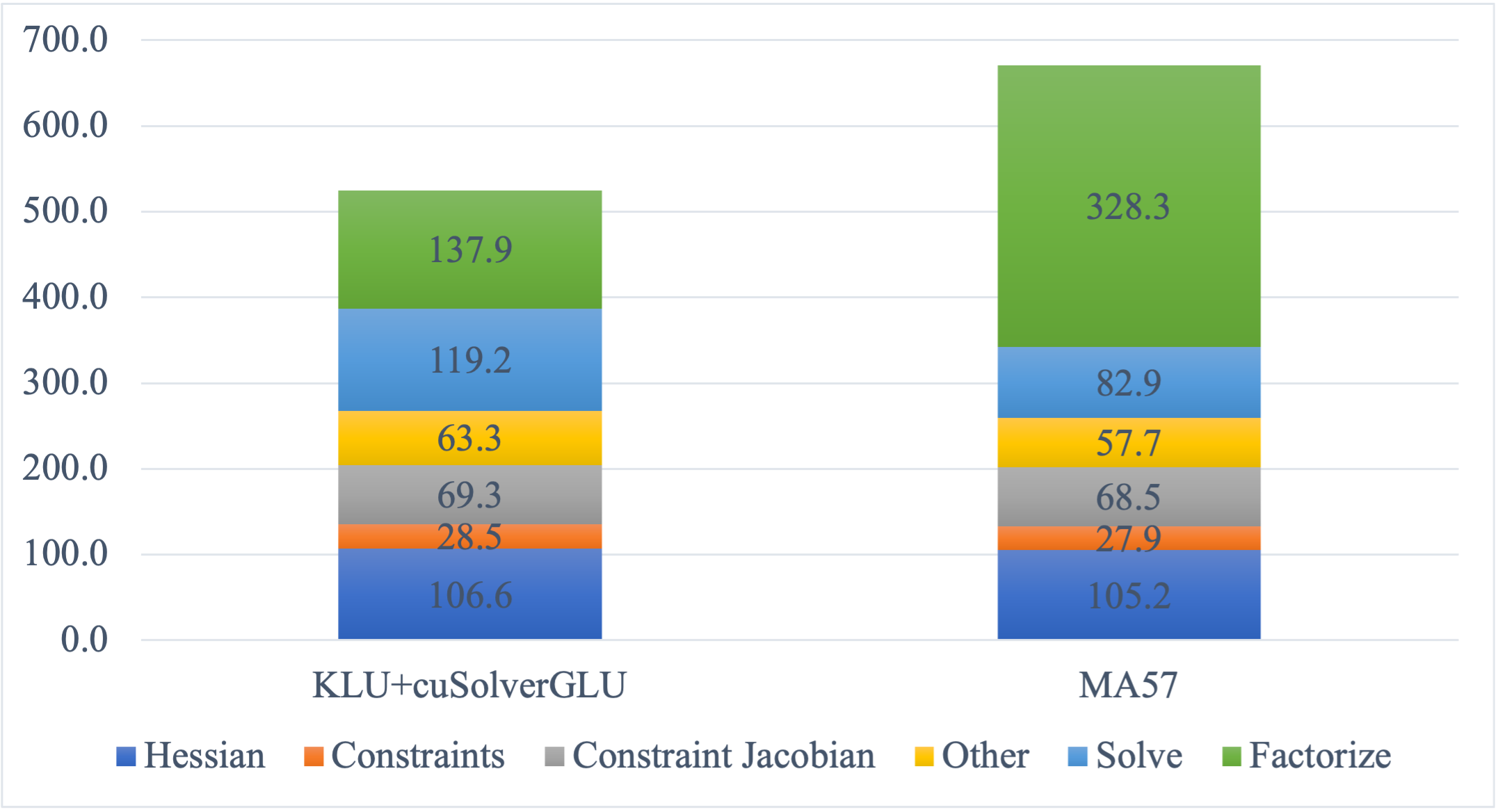}}
    \caption{Eastern U.S. grid}
  \end{subfigure}
  \caption{Comparison of the computational cost (average time per iteration in milliseconds) between \ac{cpu} and \ac{gpu} with a breakdown in terms of most expensive operations. Model evaluation time is the same in both test cases since only linear solver is GPU accelerated. 
  }
  \label{fig:ma57_vs_cusolver}
\end{figure}

Since the optimization solver typically calls the linear solver hundreds of times, the cost of one linear solve performed on \ac{cpu} can be amortized over subsequent iterations. In Fig. \ref{fig:ma57_vs_cusolver} and Table \ref{tab:ma57_vs_cusolver}, we show the computational speedup obtained using this approach. The figure shows the average cost per optimization solver iteration, including the amortized cost of one-time factorization with KLU. The numerical experiments were conducted with an IBM Power 9 \ac{cpu} and an NVIDIA V100 \ac{gpu}, which has 16 GB of high bandwidth memory and approximately 7 TF of double precision peak performance.

\begin{table}[htb]
    \centering
    \caption{Comparison of total run times for \ac{acopf} show 30\% speedup when using sparse linear solver on \ac{gpu}}
    \begin{tabular}{|p{0.3\columnwidth}|p{0.2\columnwidth}|p{0.15\columnwidth}|p{0.1\columnwidth}|}
    \hline
    \multicolumn{4}{|c|}{\textbf{Northeast U.S. grid}} \\
    \hline
     Linear solver used & KLU+cuSolver & MA57 & Speedup \\ \hline
     Total time (s) & 116.0 & 152.1 & 1.3 \\ \hline
     Number of iterations & 547 & 529 & - \\ \hline
    \multicolumn{4}{|c|}{\textbf{Eastern U.S. grid}} \\
    \hline
     Linear solver used & KLU+cuSolver & MA57 & Speedup \\ \hline
     Total time (s) & 146.7 & 196.4 & 1.3 \\ \hline
     Number of iterations & 263 & 263 & - \\ \hline
    \end{tabular}
    \label{tab:ma57_vs_cusolver}
\end{table}

The overall speedup is entirely due to acceleration of the numerical factorization. Including the cost of initial KLU factorization, we obtain an average speedup of $2.4\times$ for the factorization and $1.3\times$ overall compared to MA57 benchmark on \ac{cpu}. Somewhat surprisingly, we find that triangular solve is 30-40\% slower on \ac{gpu}.
This is likely due to a suboptimal iterative refinement embedded in \texttt{cuSolverGLU} triangular solver function. We also note there is a slight increase in ``other'' computational cost when using \ac{gpu}s. This increase is mostly due to the overhead of the CUDA API, including the cost of launching \ac{gpu} kernels. Other parts of the computation were not \ac{gpu}-accelerated and have the same cost, approximately, in both test configurations. Data movement between \ac{cpu} and \ac{gpu} during the solution process is less than 2\% of \ac{gpu} time. 

The solver requires little memory on the \ac{gpu} (1.1 GB and 2.9 GB for the 25,000-bus model of the Northeast U.S. grid and the 70,000-bus model of Eastern U.S. grid, respectively). This is in stark contrast with the implementation in \cite{abhyankar2021acopf}, which requires 21 GB of \ac{gpu} memory for the smaller 10,000-bus model of Western U.S. grid.

\section{Projected Performance Improvement with Available Technology}
\label{sec:projection}

\begin{figure}[h!]
\centering
  {\includegraphics[width=0.8\columnwidth,trim={1.1cm 0.4cm 1.1cm 7.5cm},clip]{./ma57.25k.pie}} 
  {\includegraphics[width=\columnwidth,trim={0.1cm 1cm 0.1cm 0.1cm},clip]{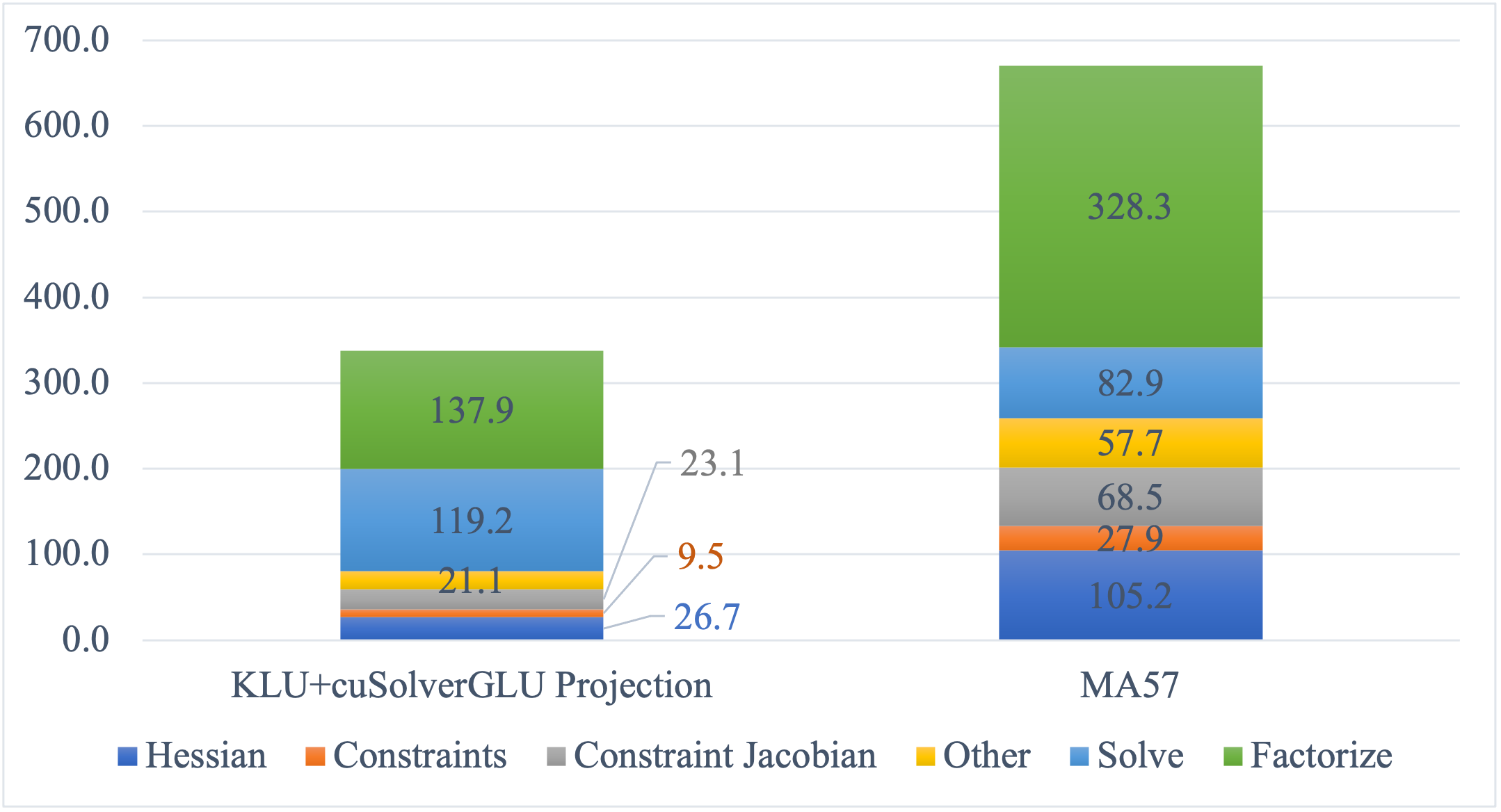}}
  \caption{A conservative projection of speedup for \ac{acopf} analysis with technology already available in \exago and \hiop libraries shows $2\times$ speedup for the Eastern U.S. grid.
  }
  \label{fig:ma57_vs_cusolver_projection}
\end{figure}

The entire performance improvement for \ac{acopf} is due to acceleration of matrix factorization on \ac{gpu} (Figure \ref{fig:ma57_vs_cusolver}).
Results in \cite{abhyankar2021acopf} show that the model evaluation could be \ac{gpu}-accelerated, as well. We can therefore make projections for further performance improvements. We conservatively anticipate a $4\times$ speedup for the sparse Hessian evaluation on \ac{gpu} and $3\times$ for constraints, constraint Jacobians and other model components. In \cite{abhyankar2021acopf}, speedup of $2$-$3\times$ was achieved on vector kernels and more than $15\times$ on dense matrix kernels for the largest grid evaluated (2,000-bus Texas grid model). While memory access is more efficient when evaluating dense matrix elements, we nevertheless believe it is safe to expect at least $4\times$ speedup for sparse matrix evaluation, especially for the large use cases we consider. Fig. \ref{fig:ma57_vs_cusolver_projection} illustrates the projected performance for the Eastern U.S. grid ($2\times$ speedup on the \ac{gpu}) if the model evaluation were performed on the \ac{gpu}.

\section{Conclusions and Next steps}
\label{sec:conslusion}


We eliminated the main bottleneck for efficient \ac{acopf} analysis on \ac{gpu} by accelerating sparse matrix factorization by $2.4\times$ compared to an MA57 baseline on \ac{cpu}. With only the linear solver accelerated, we get a 30\% overall speedup for \ac{acopf} analysis. We also show it is feasible to accelerate the entire analysis by at least $2\times$ with technology available today. 

The \ac{gpu} profiling result clearly show that the next bottleneck of the entire linear system solution pathway is now the \ac{gpu} triangular solve. 
We believe that significant improvement is possible by using a more efficient and configurable iterative refinement than the one embedded in the \texttt{cuSolverGLU} library. Triangular solves can also be approximated by Jacobi or Gauss-Seidel iteration with very promising performance results~\cite{anzt2015iterative}, but our investigation on that topic is at an early stage.

The ACOPF matrices exhibit what is known as a {\it saddle-point} structure in the literature. A specialized, \ac{gpu}-ready solver, which exploits this structure was developed in~\cite{regev2022kkt}. This solver performs triangular solve on smaller matrices, and early (standalone) results show an impressive speedup over MA57. In the future, we plan to integrate it with our software stack and evaluate its performance.

\section*{Acknowledgments}
We thank Lungsheng Chien and Doris Pan of \nvidia for their help with using the undocumented \cusolverglu module of the \cusolver library. We also thank Cosmin Petra and Nai-Yuan Chiang for their guidance when using \hiop optimization solver.
Warm thanks go to Phil Roth and Christopher Oehmen for their support of this work, and to Shaked Regev for critical reading of the manuscript and providing helpful comments.
Finally, we thank three anonymous reviewers for providing feedback that made this manuscript into a better publication.

\bibliographystyle{IEEEtran}
\bibliography{bib_pesgm_2023}

\vfill

\end{document}